\documentclass[conference]{IEEEtran}
\IEEEoverridecommandlockouts

\usepackage{cite}
\usepackage{amsmath,amssymb,amsfonts}
\usepackage{algorithmic}
\usepackage{algorithm}
\usepackage{graphicx}
\usepackage{textcomp}
\usepackage[table]{xcolor}
\usepackage{xcolor}
\def\BibTeX{{\rm B\kern-.05em{\sc i\kern-.025em b}\kern-.08em
    T\kern-.1667em\lower.7ex\hbox{E}\kern-.125emX}}
\usepackage{booktabs}
\usepackage[caption=false]{subfig}
\usepackage{lipsum}
\usepackage{tikz}

\newcommand{\RE}{\mathbb{R}}

\newcommand{\calH}{\mathcal{H}}

\usepackage{amsthm}
\newtheorem{definition}{Definition}

\theoremstyle{definition}
\newtheorem{innerexample}{Example}

\begin{document}

\title{Efficient Network Embedding by Approximate Equitable Partitions
}


\author{\IEEEauthorblockN{ Giuseppe Squillace}
\IEEEauthorblockA{\textit{IMT Lucca, Italy} \\
giuseppe.squillace@imtlucca.it}
\and
\IEEEauthorblockN{ Mirco Tribastone}
\IEEEauthorblockA{\textit{IMT Lucca, Italy} \\
mirco.tribastone@imtlucca.it}
\and
\IEEEauthorblockN{ Max Tschaikowski}
\IEEEauthorblockA{\textit{Aalborg University, Denmark} \\
tschaikowski@cs.aau.dk
}
\and
\IEEEauthorblockN{ Andrea Vandin}
\IEEEauthorblockA{\textit{Sant'Anna Pisa, Italy} \\
\textit{DTU, Denmark} \\
anvan@dtu.dk}

\thanks{This work was partially supported by Poul Due Jensen Foundation grant no. 883901, the Villum Investigator Grant S4OS, the project SERICS (PE00000014), 
the Fsc regional Tuscan project AISLEA2 J54D23000780005, 
the project Tuscany Health Ecosystem (THE), CUP: B83C22003920001, 
the project SMaRT COnSTRUCT (CUP J53C24001460006), within FAIR (CUP I53C22001380006). The last two projects are under the MUR National Recovery and Resilience Plan funded by the EU - NextGenerationEU.}
}

\maketitle

\begin{abstract}
Structural network embedding is a crucial step in enabling effective downstream tasks for complex systems that aim to project a network into a lower-dimensional space while preserving similarities among nodes. We introduce a simple and efficient embedding technique based on approximate variants of equitable partitions. The approximation consists in introducing a user-tunable tolerance parameter relaxing the otherwise strict condition for exact equitable partitions that can be hardly found in real-world networks. We exploit a relationship between equitable partitions and equivalence relations for Markov chains and ordinary differential equations to develop a partition refinement algorithm for computing an approximate equitable partition in polynomial time. We compare our method against state-of-the-art embedding techniques on benchmark networks. We report comparable---when not superior---performance for visualization, classification, and regression tasks at a cost between one and three orders of magnitude smaller using a prototype implementation, enabling the embedding of large-scale networks that could not be efficiently handled by most of the competing techniques.
\end{abstract}


\section{Introduction}

Network analysis is crucial in various fields, including, e.g., social networks \cite{freeman2004development}, bioinformatics \cite{pavlopoulos2011using}, and recommendation systems \cite{lu2012recommender}.  
The increasing availability of data leads to complex networks that are difficult to interpret and challenging computationally.  Network embedding tackles this problem by extracting  properties of a network and encoding them in a lower-dimensional space~\cite{RONE}; one  
of the main objectives is to preserve the similarity of nodes in the embedded space~\cite{cui2018survey}.

\begin{figure}[b]
\centering
\includegraphics[trim={0 1.5cm 0 0},clip,width=\columnwidth]{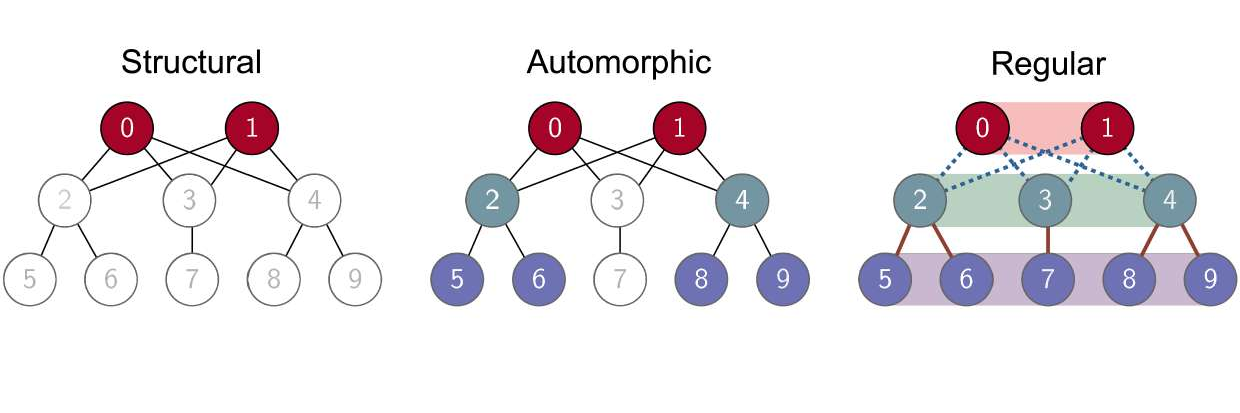}
\caption{Role equivalences (example from~\cite{axiomatic}). Nodes with the same color indicate the same role; white nodes have pairwise distinct roles.}
\label{equiv}
\end{figure}

Traditionally, the notion of similarity between nodes is based on network proximity. This approach is well suited for community detection but is not effective in capturing the \emph{role} of a node in the network~\cite{jin2021toward}, i.e., patterns of interaction beyond simple local proximity. 
The notion of role, also known as position, is well studied in the social science literature (e.g.,~\cite{merton1968social}).
Roles partition nodes into classes according to some criteria such as structural~\cite{white1983graph}, automorphic~\cite{autBorgatti}, and regular equivalence~\cite{lorrain1971structural}.
Roughly speaking, in structural equivalence, two nodes have the same role if they interact with the same set of nodes. In automorphic equivalence, two nodes are equivalent if we can define an automorphism between them. Finally, two nodes are regular equivalent if they interact with the same variety of role classes (cf. Figure~\ref{equiv}).

Structural equivalence is strict because it requires that related nodes interact with \emph{the same} set of nodes; on the other hand, automorphic equivalence suffers from well-known difficulties in its computation. Here we propose a structural embedding method for undirected networks based on equitable partitions \cite{godsil1997compact}.
This concept is also known as 
exact regular coloration \cite{everett1996exact} and graph divisor \cite{cvetkovic1980spectra}.
The idea is to assign a role in the network to each node in such a way that nodes with the same role have the same number of neighbors for each role. It is a relaxation of automorphic equivalence~\cite{axiomatic}, which can be computed efficiently in polynomial time in the size of the network~\cite{everett1996exact}.
It is stricter than regular equivalence because regular equivalence does not count the number of neighbors with the same role~\cite{white1983graph}. However, since it requires exact matching of neighbor counts, an equitable partition can still be too restrictive for real networks. 

\begin{figure}[t]
\centering
\includegraphics[width=0.7\columnwidth]{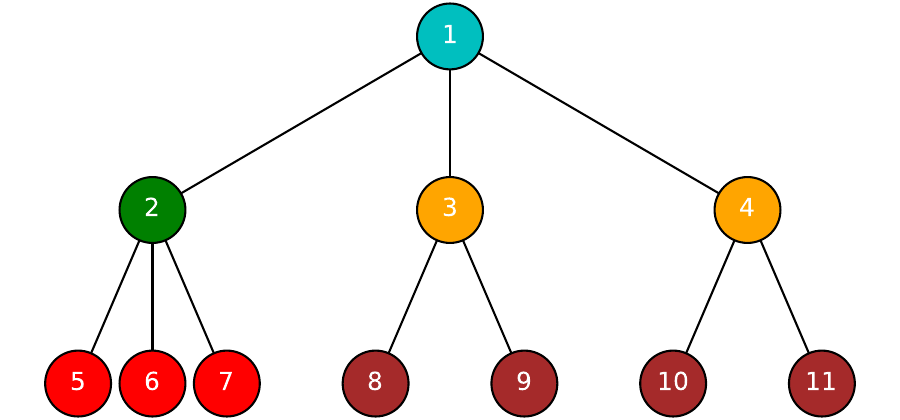}
\caption{Running example with equitable partitions.}
\label{E0}
\end{figure}

\begin{figure}[t]
\centering
\includegraphics[width=0.72\columnwidth]{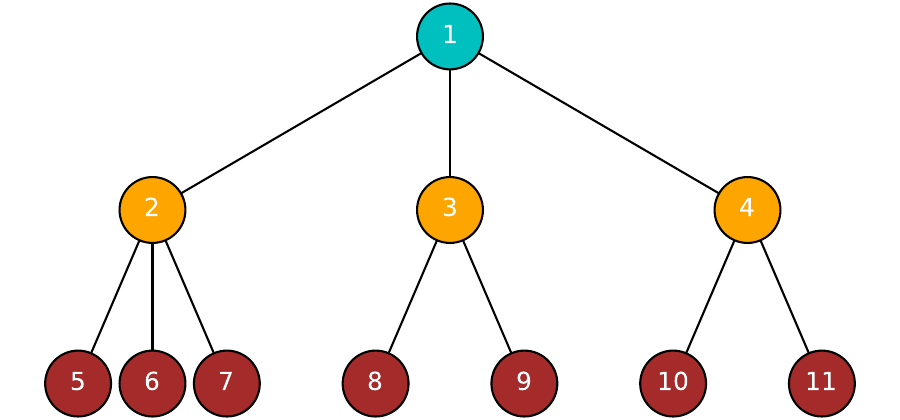}
\caption{Approximate variant of equitable partitions computed with $\varepsilon$-BE on the running example from Figure~\ref{E0} with $\varepsilon=1$.}
\label{E1}
\end{figure}
We develop a novel approach to compute a network embedding based on an approximate variant of equitable partitions that accepts differences in the number of neighbors by means of a parameter $\varepsilon$ that, informally, determines the tolerance level for identifying similar nodes.  We anticipate the results of our method on a simple network depicted in Figures~\ref{E0} and~\ref{E1}. Figure~\ref{E0} shows the equitable partition where node 2 is different from nodes~3 and~4 because it has a different degree. Our approximate variant, with $\varepsilon = 1$ accepts differences up to one node: the result is shown in Figure~\ref{E1}, where nodes 2, 3, and 4 (as well as their neighbors in the bottom layer) are identified to be approximately equivalent, thus capturing their intuitive structural similarity in the network.  

Our starting point is the observation that, for undirected networks, the equitable partition definition corresponds to an equivalence relation known as backward equivalence (BE)~\cite{PNAScttv}. BE extends exact lumpability~\cite{buchholz1994exact}, an aggregation method for Markov chains based on properties of its generator matrix, to arbitrary adjacency matrices~\cite{POPL} and nonlinear models~\cite{DBLP:conf/gecco/TognazziTTV17,gast2019size,Bioinf21}. Our network embedding method, which we call $\varepsilon$-BE, relaxes the equivalence criterion by means of the tolerance parameter $\varepsilon$. 

A crucial advantage of BE, inherited from results available for Markov chains, is the availability of an efficient algorithm for computing it that runs in $O(m \log n)$ time~\cite{Valmari}, where $m$ is the number of edges and $n$ is the number of nodes. The algorithm is based on partition refinement~\cite{paige1987three}: given an initial partition of nodes, where each block represents a candidate role, it iteratively splits blocks until the BE criterion is met. In this paper, we develop an analogous algorithm for $\varepsilon$-BE, still based on~\cite{Valmari}.
Exploiting the partition refinement nature of the algorithm, we use it as an inner step of an iterative scheme that computes an $\varepsilon$-BE partition with increasing values of $\varepsilon$, keeping track of the positions discovered at previous iterations. As will be explained later in the paper, this avoids the degenerate cases where the algorithm may aggregate too little or too much
when using a single-shot strategy with small or large enough, respectively,  values of $\varepsilon$.

The resulting partition is employed to build the network embedding $E$.
We are interested in producing a network embedding $E \in \mathbb{R}^{n\times d}$ where $n$ is the number of the nodes, and $d$ is the size of the embedding. 
This can be built by summing up the edges toward equivalent classes for each node.
Table~\ref{EmbTable} shows the embeddings for the two partitions of Figures~\ref{E0} and~\ref{E1}.
The dimension of the embedding equals the number of different blocks in each partition. 
In particular, nodes with the same role exhibit similar vectors in the network embedding and the  $\varepsilon$-BE condition can be checked by computing the difference of the embedding vectors.
For example, consider the nodes $2$ and $3$ of Table \ref{EmbTable} (right).
The difference of their embedding vectors is [0,0,1].
Each entry of this vector is $\leq \varepsilon$, and, for this reason, they present the same color in Figure \ref{E1}.

To compare $\varepsilon$-BE against state of the art, we consider the best-performing methods to network embedding covering all three aforementioned categories presented in the recent review~\cite{RONE}.  
We compare the network embeddings across various machine-learning tasks such as classification, visualization, and regression using real-world networks with up to $10^5$ nodes.

Overall, we show that $\varepsilon$-BE is able to provide competitive results, outperforming the other methods in many cases, with runtimes at least one order of magnitude faster. For larger networks, in particular, $\varepsilon$-BE provided node embeddings in a few minutes while most of the other methods timed out after 3 hours, thus promoting $\varepsilon$-BE  as a valid alternative for the structural network embedding 

\begin{table}
\resizebox{0.46\columnwidth}{!}{
\begin{tabular}[t]{c c}
     \toprule
     \multicolumn{1}{c}{Nodes} & \multicolumn{1}{c}{Embedding} \\
     \midrule
     \{1\} & [0,1,2,0,0] \\
     \{2\} & [1,0,0,3,0] \\
     \{3,4\} & [1,0,0,0,2] \\
     \{5,...,7\} & [0,1,0,0,0] \\
     \{8,...,11\} & [0,0,1,0,0]\\ 
     \bottomrule
\end{tabular}
}
\hfill
\resizebox{0.45\columnwidth}{!}{
\begin{tabular}[t]{c c}
     \toprule
     \multicolumn{1}{c}{Nodes} & \multicolumn{1}{c}{Embedding} \\
     \midrule
     \{1\} & [0,3,0] \\
     \{2\} & [1,0,3] \\
     \{3,4\} & [1,0,2] \\
     \{5,...,11\} & [0,1,0] \\
     \bottomrule
\end{tabular}
}
\vspace*{3mm}
\caption{Node representation in the embedding space associated with the equitable partition of Figure \ref{E0} (left) and Figure \ref{E1} (right). }
\end{table}\label{EmbTable}


\paragraph*{\textbf{Related work}} 
Methods for structural network embeddings can be categorized into matrix factorization, random walk, and deep learning methods.

Matrix factorization methods capture structural properties by decomposing the adjacency matrix or other related matrices into lower-dimensional matrices.
The most closely related approach to ours is RID$\varepsilon$Rs~\cite{gupte2017role}. It is also based on a notion of $\varepsilon$-equitable refinement. However, the embedding method is substantially different. Indeed, instead of the iterative scheme proposed in this paper, RID$\varepsilon$Rs computes an embedding $E_{\varepsilon}$ for each value of $\varepsilon$ ranging from 0 to the average degree of the network $d_{avg}$.
Since the overall dimension of the embedding can be large, a feature pruning technique is employed to reduce this dimension.
Subsequently, RID$\varepsilon$Rs aggregates the pruned embedding into a single matrix $E = [E_0|...|E_{d_{avg}}]$ and employs negative matrix factorization (NMF~\cite{lee1999learning}) on $E$ to generate the final embedding. 
RolX~\cite{henderson2012rolx} factorizes the network into a role membership matrix and a role-feature matrix. The matrix factorization is optimized by minimizing the reconstruction error.  GLRD extends RolX by adding further constraints in the previous optimization problem~\cite{gilpin2013guided}. GraphWave is based on a physics interpretation~\cite{GW}. It considers the application of the heat diffusion process on the network. Then it computes the graph diffusion kernel and gets the embedding using the characteristic function.
xNetMF computes a similarity matrix computed between nodes,  and then it factorizes this matrix~\cite{heimann2018regal}.
SEGK  uses graph kernels to compute node structural similarities~\cite{SEGK}. The standard one is the shortest path kernel, which uses the Nystrom  method~\cite{drineas2005approximating} for factorization.
SPaE provides a structural embedding via the Laplacian eigenmaps method~\cite{pei2019joint}.
Approaches based on random walks explore the networks with paths of a given length. According to the taxonomy in~\cite{RONE}, network embeddings are usually learned using methods from natural language processing.
One popular technique is struc2vec~\cite{S2V}. It generates a multi-layer weighted context graph, and then it employs SkipGram \cite{mikolov2013distributed} to learn the embedding. Role2vec assigns to each node a role and then builds a random walk where the generated sequence is replaced with role indicators~\cite{ahmed2020role}.
NODE2BITS employs temporal random walks where the edges are sampled with non-decreasing timestamps~\cite{jin2019node2bits}.
The last category learns the embedding through deep learning techniques such as neural networks and graph neural networks.
Among these, we mention DRNE~\cite{DRNE}.
It produces a network embedding based on the regular equivalence definition. The method is based on the training of a Long Short-Term Memory (LSTM) neural network.
GAS employs GNNs exploiting the Weisfeiler-Lehman test~\cite{guo2020role}. Other works in this direction are GraLSP~\cite{jin2020gralsp}, which incorporates random walks in the GNN, and GCC~\cite{qiu2020gcc}, which uses a pre-trained Graph Neural Network model.



\section{Background}

We consider an undirected graph with a set of nodes $V$ denoted by a symmetric adjacency matrix $A = (a_{ij}) \in \mathbb{R}^{n \times n}$, where $a_{ij} = 1$ if there is an edge between  node~$i$ and~$j$. 
Node embedding corresponds to finding a matrix $E \in \mathbb{R}^{n \times d}$ where $d < n$.
We are interested in an embedding that preserves the similarities between nodes with similar roles in the network. 

We present the definition of BE. Originally developed for hypergraphs~\cite{PNAScttv}, 
here we provide a simplified version for graphs. 
Informally, a BE is a partition of the nodes such that nodes belonging to the same block present the same cumulative degrees towards equivalence classes.  

\begin{definition}[Backward equivalence, simplified to graphs  from~\cite{PNAScttv}]\label{def:be}
For an adjacency matrix $A \in \RE^{n \times n}$, a partition of the nodes $\mathcal{H}$  is called a backward equivalence (BE) when 
\[
\sum_{k \in B'} a_{ki} = 
\sum_{k \in B'} a_{kj}\]
for all blocks $B,B' \in \calH$ and $i,j \in B$.
\end{definition}

When simplified to graphs, BE is equivalent to the equitable partition definition presented in \cite{gupte2017role}, as is apparent from comparing the previous and the forthcoming definitions. 
\begin{definition}[Equitable partition, adapted from~\cite{gupte2017role}]\label{def:eqpar_andrea}
	For a network with nodes $V$,  and  edges $E$, 
	a partition $\mathcal{H}$ of   $V$ is an equitable partition if and only if
	\[deg(v_i,B')=deg(v_j,B') \]
	for all blocks $B,B' \in \calH$, and $v_i,v_j \in B$, where  \[deg(v_l,B'') = |\{v_k | (v_l,v_k) \in E \text{ and } v_k \in B'' \}|\]
\end{definition}

The colors in Figure \ref{E0} identify the different blocks of a BE/equitable partition.
We now relax the notion of BE to account for tolerances in node degrees. Here, we consider the tolerance $\varepsilon$ as a measure to accept approximate and not only exact equitable partitions.

\begin{definition}[$\varepsilon$-BE]\label{def:epsBE}
For an adjacency matrix $A \in \RE^{n \times n}$, a partition of the nodes $\calH$ is called $\varepsilon$-BE partition when
\[
 \left|\sum_{k \in B'} a_{ki} - \sum_{k \in B'} a_{kj} \right| \leq \varepsilon\]
 for all blocks $B,B' \in \cal H$ and $i,j \in B$.
 \end{definition}

We can see that $\varepsilon$-BE corresponds to BE for $\varepsilon=0$. 
However, by increasing $\varepsilon$, two nodes can be related even if they have similar degrees towards (approximately) equivalent blocks. The parameter $\varepsilon$  tunes the level of similarity of the nodes in the same block. 
For instance, setting $\varepsilon = 1$ allows larger identifications of similar states in Figure~\ref{E1} as opposed to the equitable partition in Figure~\ref{E0}. 

This definition is inspired by a technique to reduce dynamical systems called $\varepsilon$-BDE \cite{JLAMP}.
The main difference between the two approaches lies in the use of the tolerance $\varepsilon$.
In $\varepsilon$-BDE, the cumulative difference in the degree towards all equivalent blocks should be less than $\varepsilon$.
Instead $\varepsilon$-BE is less strict as it requires the difference towards each block to be smaller than $\varepsilon$.
With this change, it turns out that, for undirected networks, $\varepsilon$-BE  corresponds to the notion of approximate equitable partition from \cite{gupte2017role}.

We build a node embedding using $\varepsilon$-BE as follows.
\begin{definition}[$\varepsilon$-BE embedding] \label{emb}
Let us consider an adjacency matrix $A \in \RE^{n \times n}$ and an $\varepsilon$-BE partition $\mathcal{H} = \{ B_1,...,B_d \}$ with $d < n$.
We define the network embedding matrix $E \in \RE^{n \times d}$ as follows:
\begin{equation}
    E_{ik} = \sum_{j \in B_{k}}  a_{ij}
\end{equation}
for all i = \{1,...,n\} and
                k = \{1,...,d\}.
\end{definition}
The matrix $E$ collects the encoding of each node in the low dimensional space $\RE^d$. The value $E_{ik}$ corresponds to the number of edges connecting node $i$ to nodes in the equivalence class $B_k$.
Notice that, for $\varepsilon = 0$,  BE-equivalent nodes have the same vector in the embedded space $E$.

\begin{figure*}[t]
\centering
\includegraphics[width=\linewidth]
{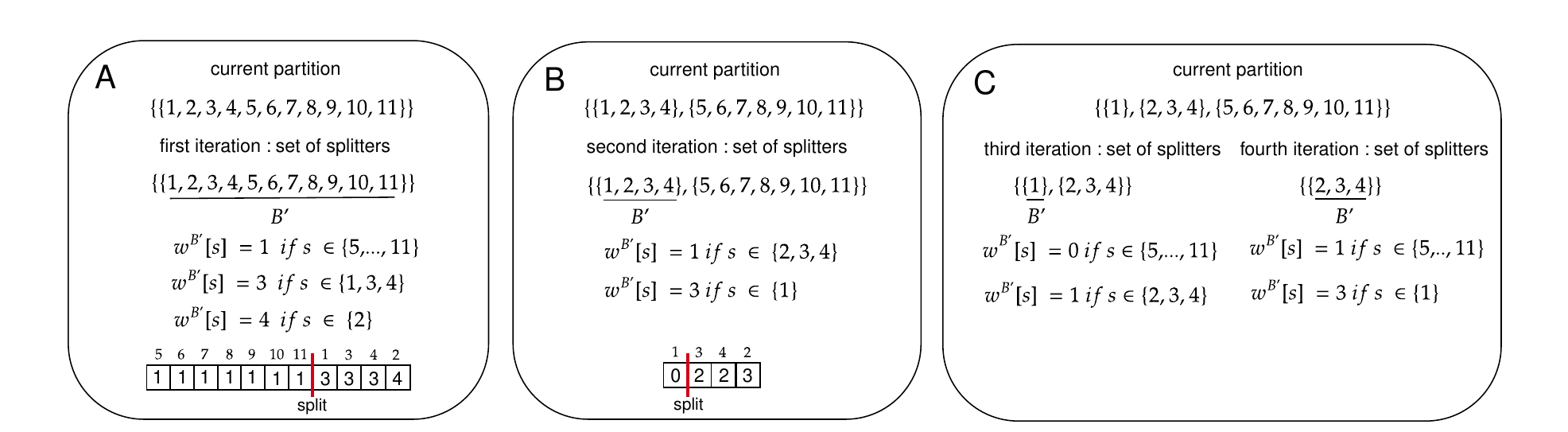}
\caption{
Execution of Algorithm 1. We indicate the current splitter by underlining it. The weight $w^{B'}$ denotes the number of edges connected to elements of the splitter $B'$. The arrays at the bottom show the splitting phase of a certain block $B$ with respect to the current splitter. The red line indicates where the block is split because it does not respect the tolerance $\varepsilon$.}
\label{diagram}
\end{figure*}

\section{Computation of Iterative $\varepsilon$-BE}

\subsection{Partition-refinement Algorithm}

We present an efficient algorithm to find an $\varepsilon$-BE partition based on the Markov chain lumping algorithm from~\cite{Valmari}.
We report the pseudocode in Algorithm~\ref{Valm}.
The algorithm starts with an initial partition and refines it, splitting the blocks, until it gets a partition where the $\varepsilon$-BE condition is satisfied.  
In general, the initial partition can be used to avoid the aggregation of undesired nodes; indeed, nodes in different blocks of the initial partition cannot be aggregated. 
Throughout this paper, initial partitions always consist of a single block containing all nodes. We will see how the notion of initial partition is instrumental to obtain an iterative version of the algorithm in Section~\ref{sec:epsBE}.

\begin{algorithm}[t!]
\caption{ $\varepsilon$-BE}
 	\begin{algorithmic}[1]
	 \REQUIRE Adjacency matrix $A$, initial partition $\calH_{in}$, tolerance $\varepsilon \geq 0$.
    \STATE $U = \mathcal{H}_{in}; \;$ $B_T = \{\}; \;$ \label{row0}
    \WHILE{$U \neq \emptyset$ }
    \STATE $w^{B'}[s] = 0 $ for each node $s$ \label{row1}
    \STATE $B' = $ remove a block from $U$. 
    \STATE $S_T = \{\}$\label{row2}
    \FOR{$s' \in B'$}\label{row3}
        \FOR{$s$ connected to $s'$}
            \IF{$w^{B'}[s] == 0$}
                \STATE $S_T = S_T \cup \{s\} \;$ $w^{B'}[s] = A[s][s']$
            \ELSE
                \STATE $w^{B'}[s] = w^{B'}[s] + A[s][s']$
            \ENDIF
        \ENDFOR
    \ENDFOR\label{row4}
    \FOR{$s \in S_T$}\label{row5}
        \STATE $B$ = the block that contains $s$
        \IF{$B$ not in $B_T$}
            \STATE $B_T = B_T \cup \{ B \}$
        \ENDIF
    \ENDFOR\label{row6}
    \WHILE{$B_T \neq \emptyset$ }\label{row7}
        \STATE $B =$ remove a block from $B_T$
        \STATE find possible majority candidate of $w^{B'}[s]$ for $s \in B$\label{pmcrow}
        \STATE sort $B$ without nodes with $w^{B'}[s] = pmc$ \label{sortrow}
        \STATE add nodes with $w^{B'}[s] = pmc$ to $B$
        \STATE partition $B$ in $B_1,B_2,...,B_l$ respect to tolerance $\varepsilon$
        \IF{$l > 1$} \label{rowcond}
            \STATE add $B_1,B_2,...,B_l$ to $U$
        \ENDIF
    \ENDWHILE \label{row8}
    \FOR{$s \in S_T$ }
    \STATE $w^{B'}[s] = 0$
    \ENDFOR
    \ENDWHILE \label{li:initend}
	\end{algorithmic}
    \label{Valm}
\end{algorithm}

The algorithm iteratively scans a list of \emph{splitters}, i.e., partition blocks. The list is initialized with the initial partition. It removes each splitter from the list and checks whether all blocks satisfy the condition for the splitter. The blocks not satisfying the condition are refined such that the $\varepsilon$-BE condition is satisfied for the splitter; the newly identified blocks are added to the list of splitters. The computation terminates when the list of splitters is empty.

For an illustrative explanation, Figure~\ref{diagram} reports the computation of $\varepsilon$-BE for the running example from Figure \ref{E0} with $\varepsilon = 1$.
It is divided into three phases (panels A, B, and C in Figure~\ref{diagram}).
In panel A, we show the set of splitters $U$ as the blocks in the initial partition (line~\ref{row0}).
The procedure removes a block from $U$ and initializes $S_T$, the set of nodes connected to any node in the current splitter (lines \ref{row1}-\ref{row2}).
In Figure \ref{diagram} we indicate the current splitter with $B'$.
Then, the algorithm computes the $w^{B'}[s]$ as the number of edges connecting the node $s$ to a node in the current splitter $B'$ (lines \ref{row3}-\ref{row4}).
In panel A, $w^{B'}[s]$ is computed for the first block. 
The nodes $\{5,...,11\}$ are connected to one node belonging to $B'$, $\{1,3,4\}$ to three nodes, and $\{2\}$ to four nodes.

The blocks containing at least a node connected to the current splitter are retrieved and stored in $B_T$ (lines \ref{row5}-\ref{row6}).
For each block $B$ in $B_T$, the algorithm sorts $w^{B'}[s]$ with $s \in B$ and splits them into blocks (lines \ref{row7}-\ref{row8}).
Here, differently than \cite{Valmari}, we allow a tolerance $\varepsilon$ in the aggregation of the nodes.

Figure~\ref{diagram} shows the splitting process in the bottom part of panel A.
The nodes $s$ belonging to $B$ are sorted by $w^{B'}[s]$. 
If the difference between the maximum and minimum nodes is greater than $\varepsilon$, it splits the block.
A red line indicates the split. For instance, when node $\{1\}$ is considered, the difference is greater than $\varepsilon = 1$ and it splits the block.

In line \ref{pmcrow}, the algorithm finds a possible majority candidate (\texttt{pmc}) in the array $w^{B'}[s]$ with $s \in B$. That is, a value $x$ that is present in more than half of the positions of the array; in the case where no such value exists, the \texttt{pmc} can be any value present in the array. 
The \texttt{pmc} can be computed in linear time and is useful to reduce the number of nodes to be sorted in line \ref{sortrow}.
Here the algorithm sorts the nodes in $B$ except the ones with $w^{B'}[s] = pmc$. For a detailed discussion of the algorithm, see~\cite{Valmari}.

After the splitting process, in panel B of Figure~\ref{diagram}, the current partition is composed of two blocks. The first block $\{1,...,4\}$, where the difference is 0, and the second block $\{5,...,11\}$, where the difference is 1.
On line \ref{rowcond}, we evaluate whether the number of new blocks exceeds 1. If this condition holds true, the new blocks are incorporated into $U$ as additional splitters. When $l$ equals 1, $B$ is not split into any new blocks, so we avoid adding itself to $U$.

The algorithm considers a new splitter $B'$ and tries to split the first block.
It considers the sorted $w^{B'}$ and it finds that node $1$ can be separated.
At this point, the set of splitters and the partition will be updated.
In phase C, we consider the remaining splitters. No actual refinements are performed, hence the procedure terminates. Therefore, the output is a partition with three blocks as pictured in Figure~\ref{E1}.

\begin{figure*}[t]
\centering
\includegraphics[width=0.90\linewidth]{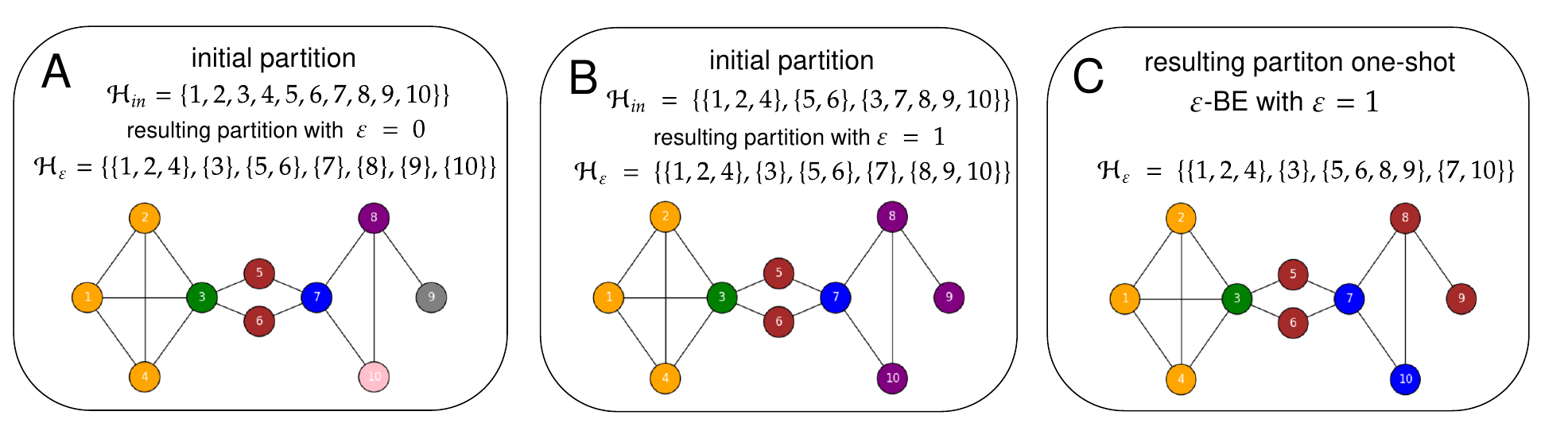}
\caption{ Execution of the iterative $\varepsilon$-BE on a network composed of a complete clique on the left, and another on the right with missing edges. We show the first and the second iteration with $\varepsilon$ equal to 0 and 1 in panels A and B. For each panel, we indicate the initial partition and the resulting one after the application of $\varepsilon$-BE.
The resulting partition of a phase will be employed as the initial partition in the following phase after merging the singleton blocks. 
The initial partition $\mathcal{H}_{in}$ in panel B corresponds to $\mathcal{H}_{\varepsilon}$ in panel A where the singleton blocks are merged into a single block.
We depict the network with the different colors assigned by the resulting partition. Panel C represents the result of the application of Algorithm \ref{Valm} without the iterative scheme. }
\label{execIT}
\end{figure*}

\begin{figure*}[h!t]
\centering
\includegraphics[width=0.70\linewidth]{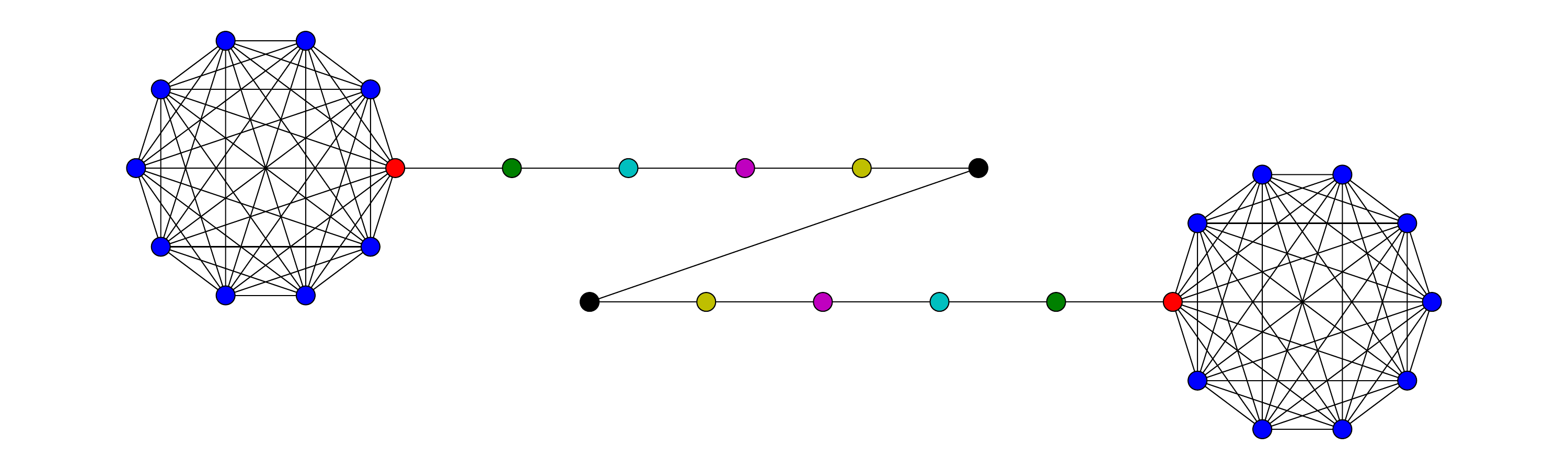}
\caption{Barbell role assignment.}
\label{barbell}
\end{figure*}

\begin{algorithm}[t]
\caption{Iterative $\varepsilon$-BE.}\label{itBE}

\begin{algorithmic}[1]
	 \REQUIRE Adjacency matrix $A$, initial partition of nodes $\mathcal{H}_{in}$, an initial tolerance $\varepsilon_0$,  step size $\delta \ge 0$,  maximum tolerance $\Delta \ge 0$.

  \STATE $\varepsilon$ = $\varepsilon_0$
	
  \WHILE{$\varepsilon$ $\leq$ $\Delta$}\label{itrow1}

    \STATE $\mathcal{H}_{\varepsilon}$ = $\varepsilon$-BE$(A,\mathcal{H}_{in},\varepsilon)$\label{itrowBE}


      \STATE $\mathcal{H}_{in}$ =  joinSingletons($\mathcal{H}_{\varepsilon}$) \label{itrowsingl}



	 \STATE $\varepsilon$ = $\varepsilon + \delta$ \label{itrowinc}

	
  \ENDWHILE \label{itrow2}
	
  \RETURN $\mathcal{H}_{\varepsilon}$
  \end{algorithmic}
\end{algorithm}

\subsection{Iterative $\varepsilon$-BE}\label{sec:epsBE}

One of the crucial aspects of Algorithm~1 is the choice of the tolerance $\varepsilon$: increasing this value leads to partition with few, large blocks, but it may collapse the network too aggressively. For instance, running it with $\varepsilon = 3$ on the running example in Figure \ref{E0} results in the trivial partition composed of a single block.
For an effective network embedding, we propose an iterative strategy. The pseudocode is shown in Algorithm~\ref{itBE}.

It requires the following parameters: an initial tolerance, denoted by $\varepsilon_0$, a step size $\delta$, which determines the increment of $\varepsilon$ at each step, and $\Delta$, the maximum value of $\varepsilon$ to be considered.
The main loop iterates in lines \ref{itrow1}-\ref{itrow2} until it reaches the maximum threshold $\Delta$.
In line \ref{itrowBE}, Algorithm~\ref{Valm} is called to find the partition up to a certain threshold $\varepsilon$.
The so-computed partition is stored in $\mathcal{H}_{\varepsilon}$ (line \ref{itrowBE}). In line \ref{itrowsingl}, all the singleton blocks in the current partition $\mathcal{H}_{\varepsilon}$ are joined into one block for the next iteration. The intuition is to attempt node aggregation for smaller values of  $\varepsilon$ first. If that fails, i.e., nodes are eventually outputted as singletons blocks,  the merging of such nodes is used to attempt aggregation for the larger $\varepsilon$ values in the next iterations.
Finally, in line \ref{itrowinc}, the algorithm increments the value of $\varepsilon$ by $\delta$.

Figure~\ref{execIT} provides an illustrative example.  It shows a network with two cliques connected by edges; the one on the left side is a complete clique (nodes 1, 2, 4, 3), while the one on the right has some missing edges (nodes 7, 8, 9, 10).
Intuitively, one expects to group the elements of the complete clique with $\varepsilon = 0$, while for the incomplete one, we expect to find an association of the cliques' nodes considering a tolerance greater than 0. 
We employ the Algorithm \ref{itBE} in the example in Figure \ref{execIT}. We set up $\varepsilon_0 = 0$, $\delta = 1$ and $\Delta = 1$. 
this corresponds to two iterations, with $\varepsilon$ equal to 0 and 1, shown in panels A and B of Figure~\ref{execIT}, respectively.

Panel A shows the initial partition $\mathcal{H}_{in}$ and the coloration for $\varepsilon = 0$.
As expected, the nodes in the left clique present the same role. In panel B the initial partition $\mathcal{H}_{in}$ presents the merged singleton blocks $\{3\}$, $\{7\}$, $\{8\}$, $\{9\}$, $\{10\}$ of $\mathcal{H}_{\varepsilon}$ at the previous iteration of panel A. 

 In the subsequent iteration, the algorithm computes the partition with $\varepsilon = 1$, considering as new initial partition the one computed in the previous iteration. The results are depicted in panel B, where, due to the tolerance parameter $\varepsilon$, nodes within the incomplete clique are denoted with the same color. This example demonstrates how the algorithm can collect similar groups of nodes while gradually increasing the tolerance $\varepsilon$.

To show the advantages brought by the iterative scheme, panel C shows the output of a \emph{one-shot} application of Algorithm~\ref{Valm} directly with $\varepsilon = 1$. This comparison shows that the iterative algorithm is able to better assign the roles to nodes in the two cliques, while the one-shot application identifies nodes 5, 6, 8, and 9, in contrast with their intuitively different roles in the network.

\subsection{Complexity}

The original partition-refinement algorithm from~\cite{Valmari} runs in $O(m \log(n))$ time, where $n$ is the number of nodes and $m$ is the number of edges.
The logarithmic term is achieved by discarding the largest block in each iteration.
Unfortunately, for an approximate reduction, as presented here, discarding the largest block does not assure correctness (i.e., the partition does not satisfy the $\varepsilon$-BE condition).
Consequently, the complexity of the Algorithm \ref{Valm} is $O(mn)$. The splitting of the blocks according to the tolerance $\varepsilon$ can be implemented in linear time once the nodes are sorted for their weights $w$ and does affect the overall complexity.

Algorithm \ref{Valm} is employed as an inner step in Algorithm~\ref{itBE}. An important point to consider is how many times lines \ref{itrow1}-\ref{itrow2} are repeated.
In this case, the maximum threshold $\Delta$ and the step size $\delta$ have to be considered. 
The inner loop will be repeated $\lceil \Delta / \delta \rceil$ times. Since every element of the adjacency matrix is either 0 or 1, the minimum $\delta$ one can consider is 1, so that complexity is $O(mn\Delta)$.
Despite this, the parameter $\Delta$ is always orders of magnitude smaller than $n$ to avoid trivial results where all the nodes belong to the same block. 

\section{Numerical Experiments}\label{sec:exps}

In this section, we present the experimental results achieved by $\varepsilon$-BE compared with state-of-the-art methods for network embedding.
All the experiments and datasets with the implementation of our method are accessible on the GitHub page.\footnote{GitHub link: https://github.com/EmbNet01/EmbNetworks.git}
We considered the best methods reported in the recent review \cite{RONE}.
Specifically, we chose Graphwave~\cite{GW}, SEGK~\cite{SEGK} and RID$\varepsilon$Rs~\cite{gupte2017role} as representatives of matrix factorization techniques, struc2vec~\cite{S2V} as a random-walks-based approach, and DRNE~\cite{DRNE} and GAS~\cite{guo2020role} as representatives of deep learning methodologies.
The implementation details for each method were taken from GitHub repositories mentioned in the respective publications or in~\cite{RONE}.\footnote{GitHub repositories accessed on 01/06/2023.}
Similarly to~\cite{SEGK,RONE,GW}, the comparison was conducted on synthetic and real networks considering three machine-learning tasks: visualization, classification, and regression.
All experiments were run on a machine equipped with an Intel(R) Xeon(R) CPU E7-4830 with 500 GB RAM.
We set up a timeout of 3 hours for all the computations reported in this section.

\subsection{Set-up}
Most of the methods in network embedding allow the user to select the desired embedding dimension $d$ for the network.
However, choosing the optimal dimension is challenging~\cite{gu2021principled}. In this study, we followed~\cite{SEGK} and~\cite{GW} in adopting the default embedding dimensions recommended for each method. Specifically, we set 
$d$ to the following values: 20 (SEGK), 100 (Graphwave), 64 (DRNE), and 128 (struct2vec and GAS). Regarding RID$\varepsilon$Rs, the embedding dimension was determined through the NMF method, optimizing the minimum description length criterion. In addition to the size of the embedding, each method requires specifying other parameters.
We set up all these parameters with their default values.

$\varepsilon$-BE yields a discrete embedding based on the partition computed by Algorithm \ref{itBE}. The reduction with the maximum number of blocks corresponds to a partition where every node is a singleton, resulting in an embedding equivalent to the adjacency matrix itself. In this scenario, similar nodes are defined as those that are connected to the exact same set of nodes. Conversely, considering a sufficiently large 
$\varepsilon$, the partition with the minimum number of blocks groups all nodes into a single block, resulting in an embedding dimension 
$d$ equal to 1. Here, each entry represents the sum of the elements in the respective columns, and similar nodes are defined as those with comparable degrees. These represent two extreme reductions of the embedding space. 
Given that the primary objective of network embedding is to offer an encoding where $d$ is significantly smaller than $N$, we set the parameters to achieve a reduction of at least $70\%$ of the network dimension $N$. 

\begin{figure*}
\centering
\subfloat[$\varepsilon$-BE]{
\includegraphics[width=0.27\linewidth]
{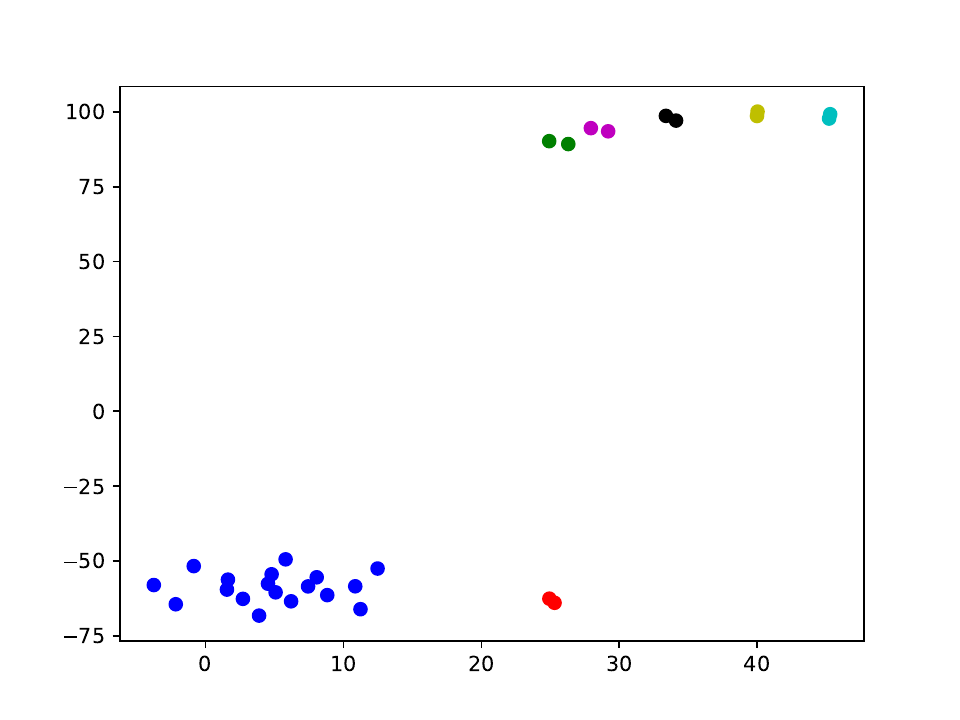}}
\subfloat[Graphwave]{\includegraphics[width=0.27\linewidth]
{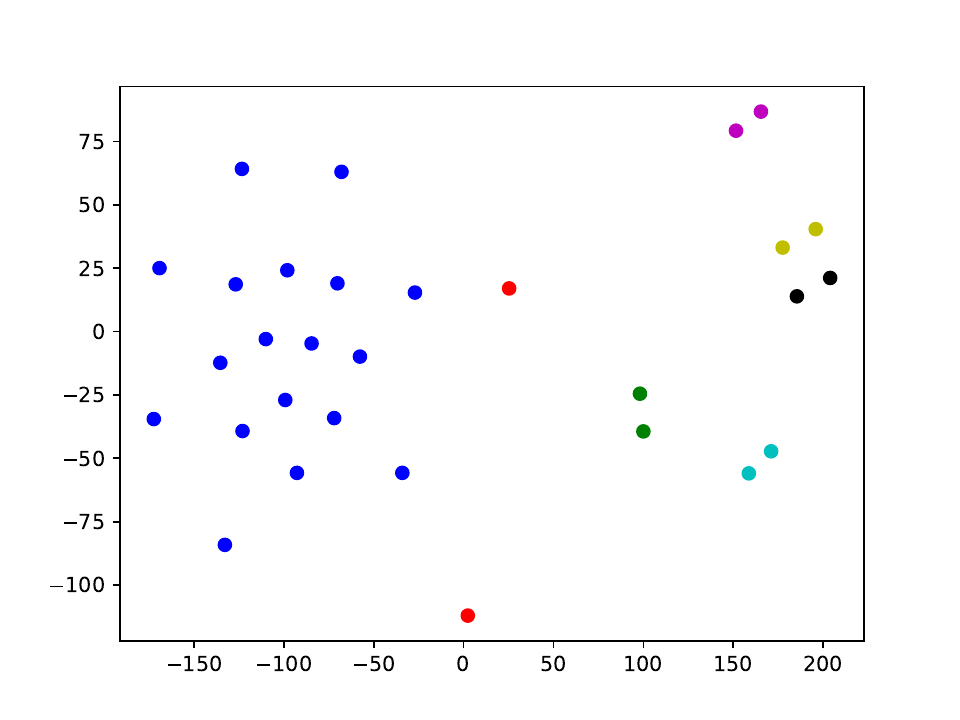}} 
\subfloat[DRNE]{\includegraphics[width=0.27\textwidth]
{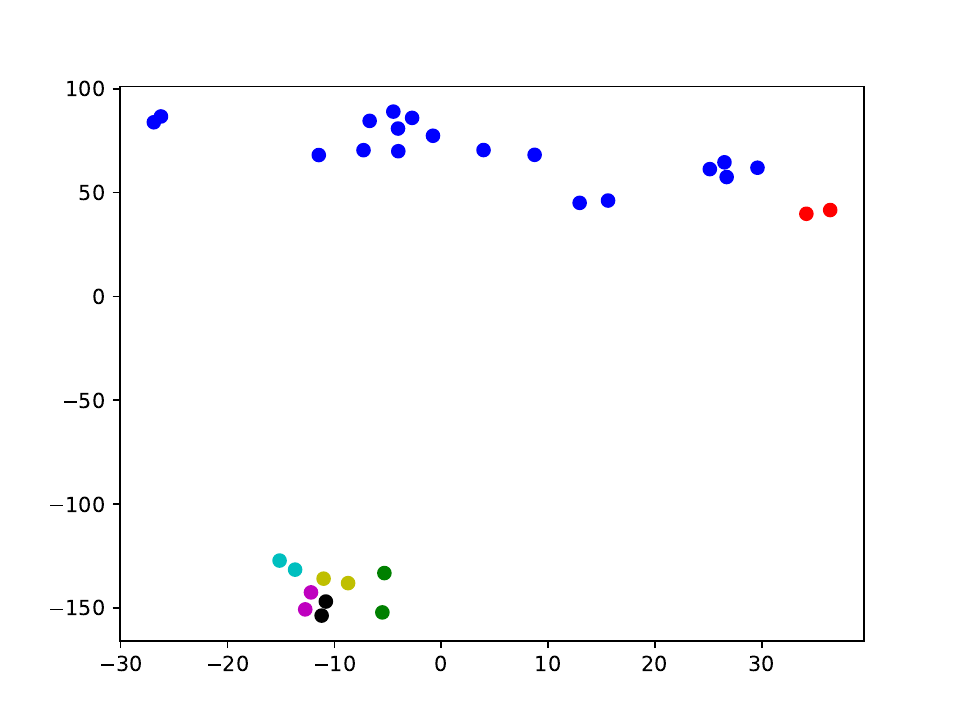}} 

\subfloat[struc2vec]{\includegraphics[width=0.27\textwidth]
{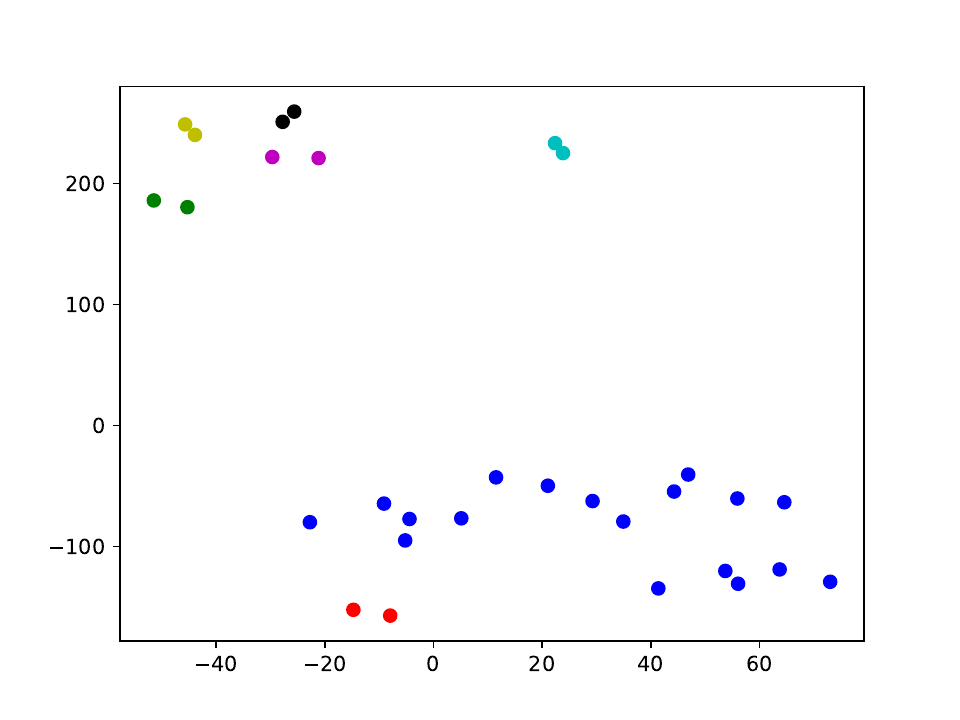}}
\hspace{-0.6cm}
\subfloat[SEGK]{\includegraphics[width=0.27\textwidth]
{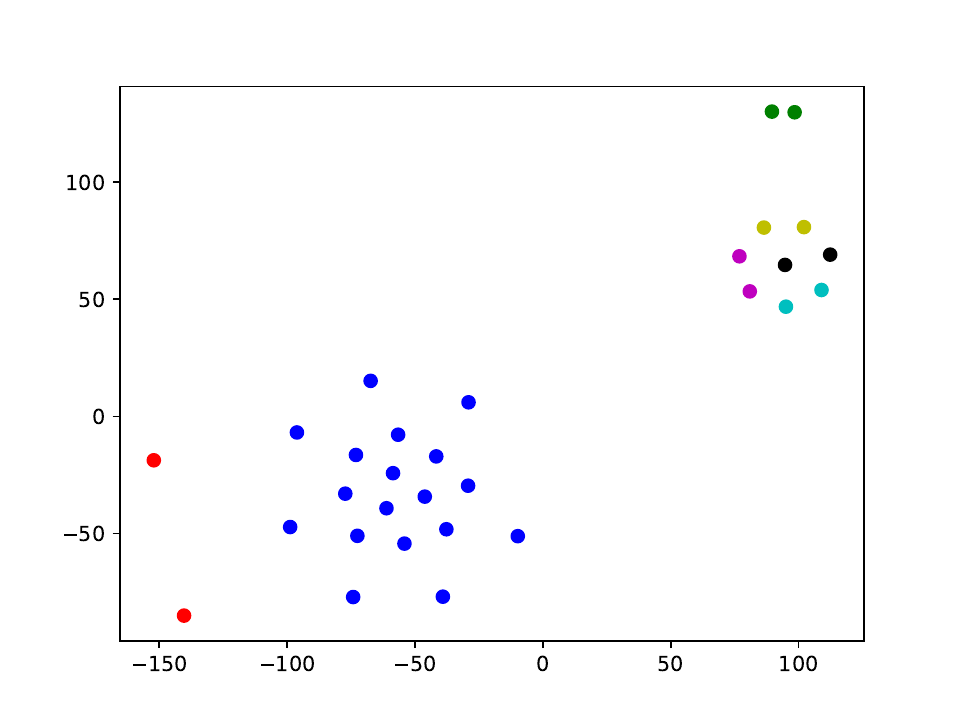}}
\hspace{-0.6cm}
\subfloat[GAS]{\includegraphics[width=0.27\textwidth]{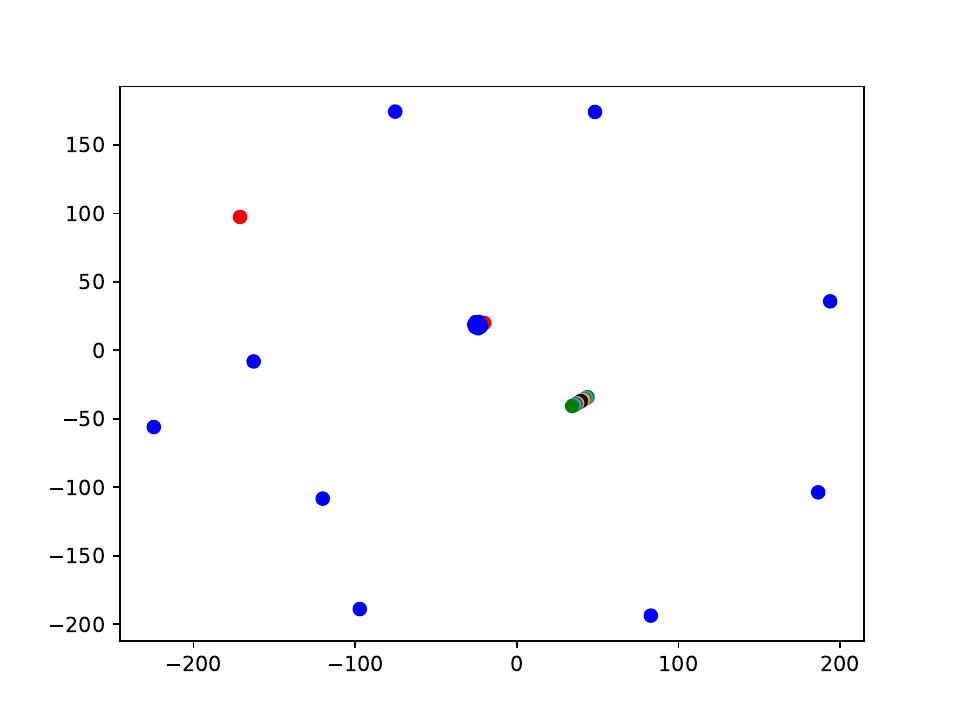}}
\hspace{-0.6cm}
\subfloat[RID$\varepsilon$Rs]{\includegraphics[width=0.27\textwidth]
{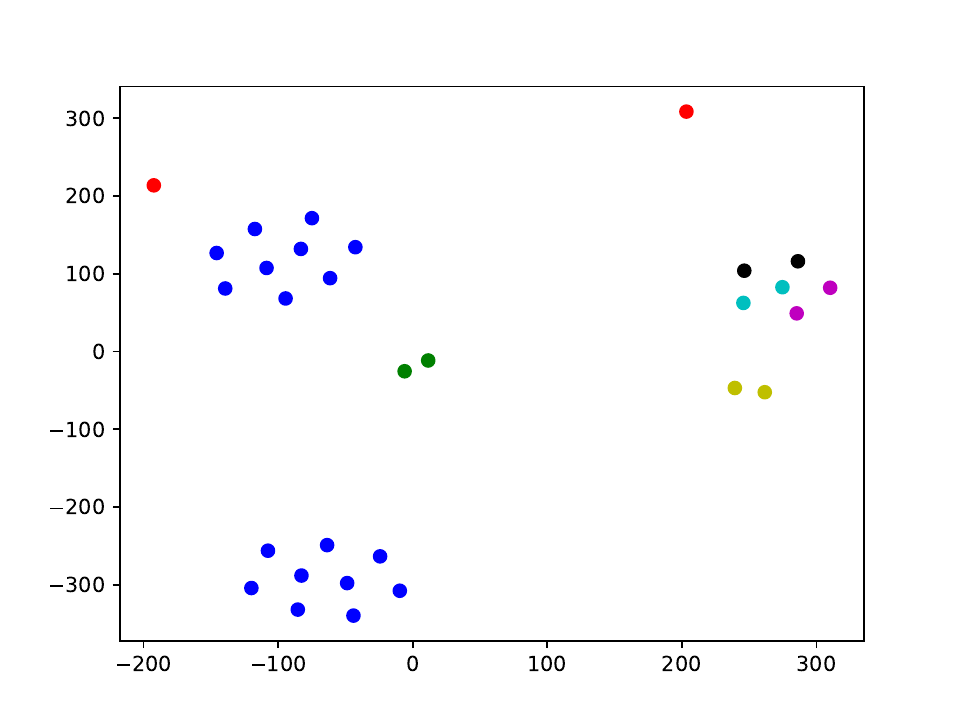}}
\caption{ Barbell visualization with t-SNE.}
\label{TSNE}
\end{figure*}

\begin{figure*}
\centering
\subfloat[$\varepsilon$-BE]{
\includegraphics[width=0.27\textwidth]
{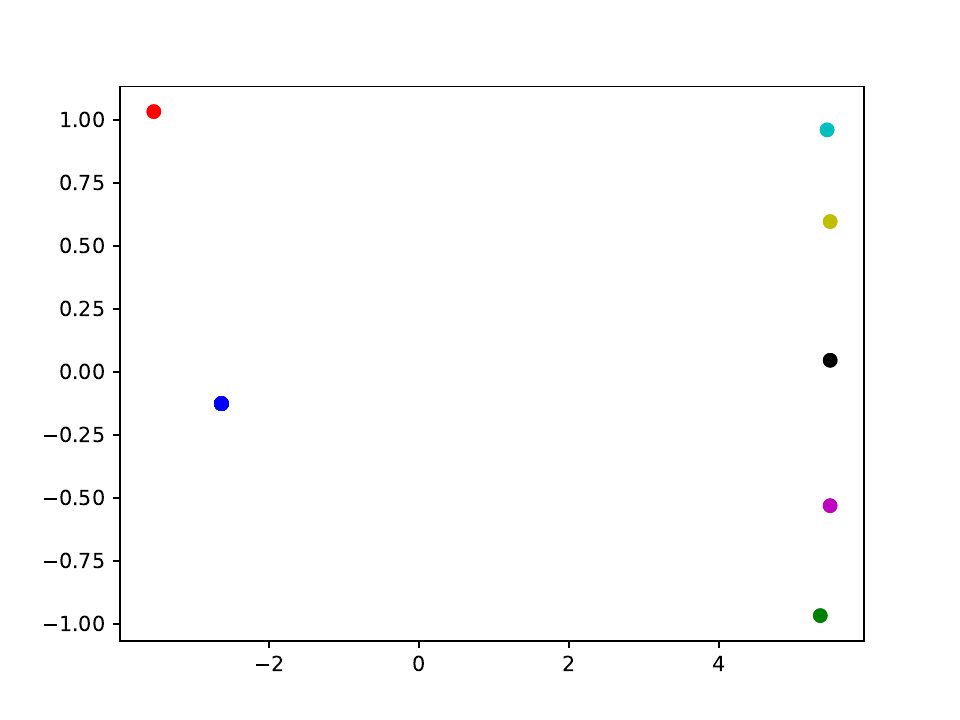}}
\subfloat[Graphwave]{\includegraphics[width=0.27\textwidth]
{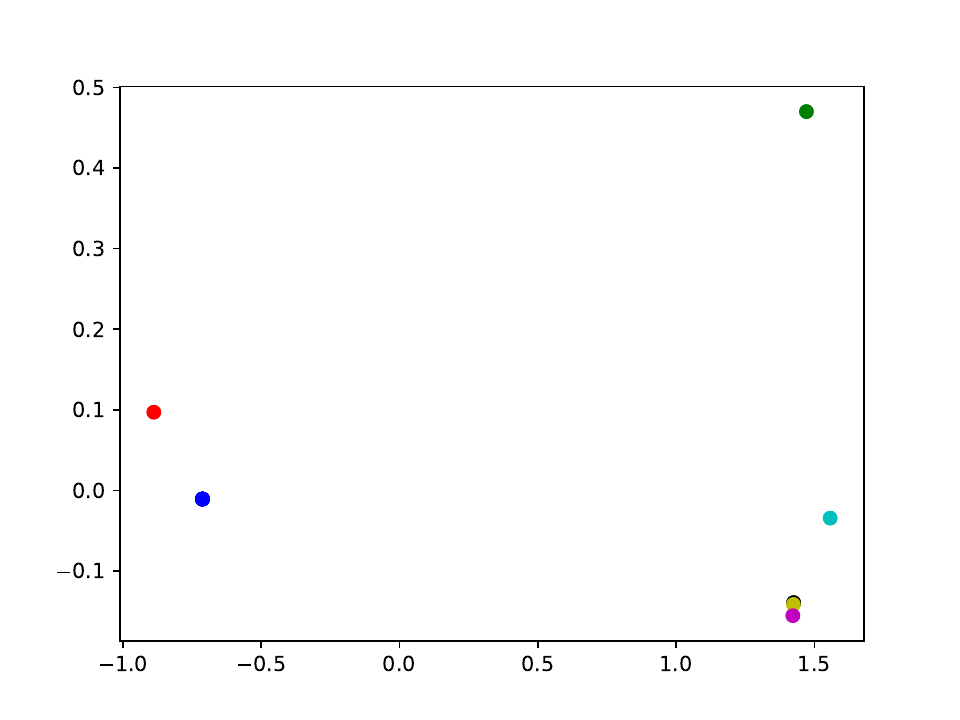}} 
\subfloat[DRNE]{\includegraphics[width=0.27\textwidth]
{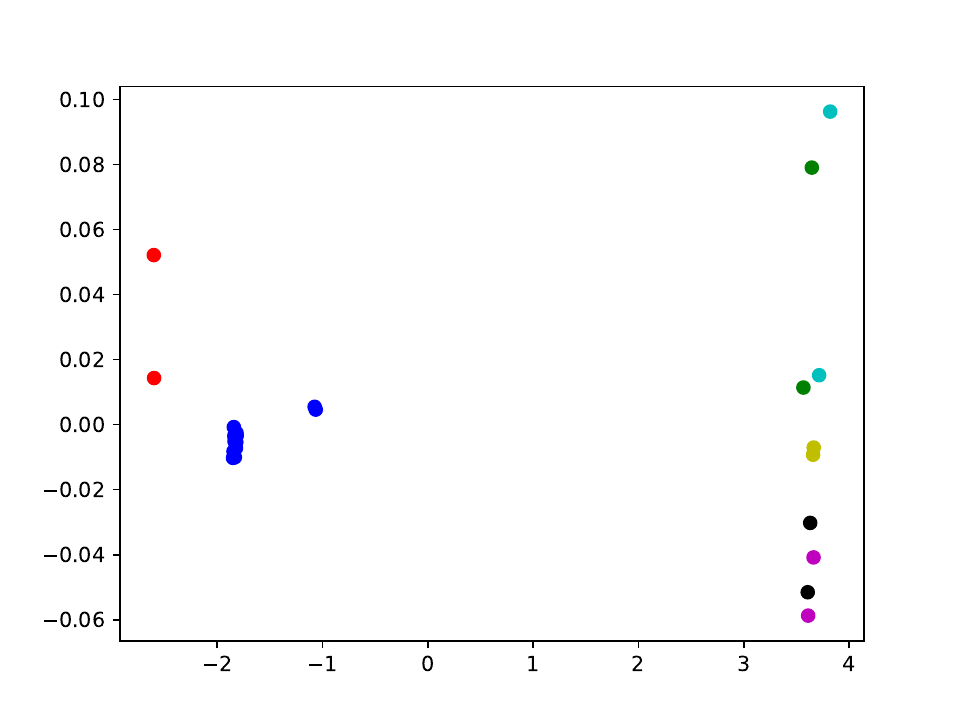}} 

\subfloat[struc2vec]{\includegraphics[width=0.27\textwidth]
{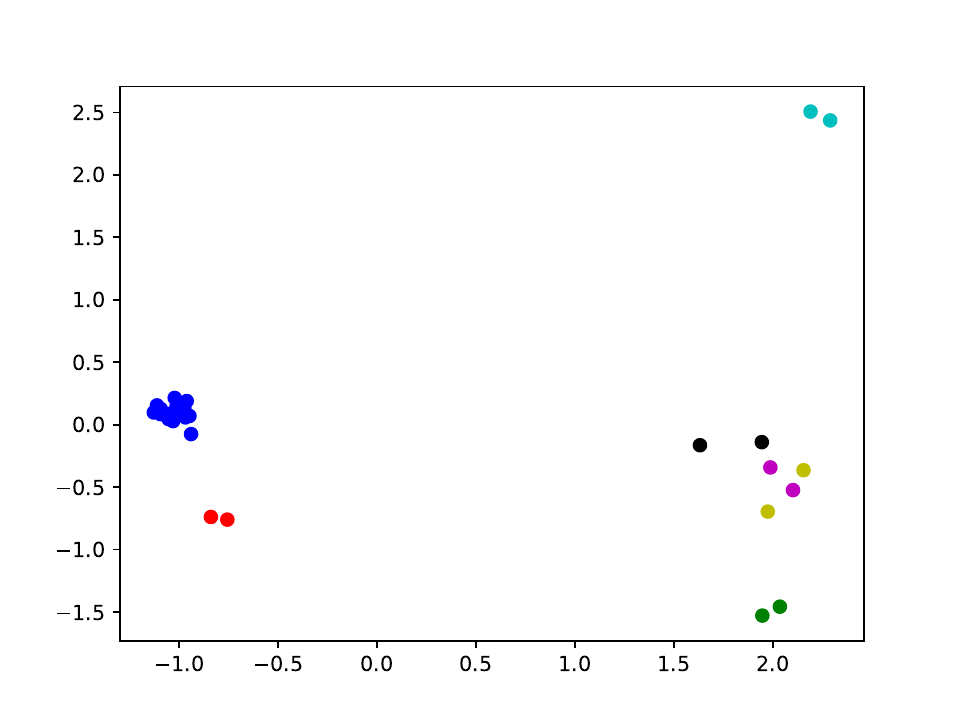}}
\hspace{-0.6cm}
\subfloat[SEGK]{\includegraphics[width=0.27\textwidth]
{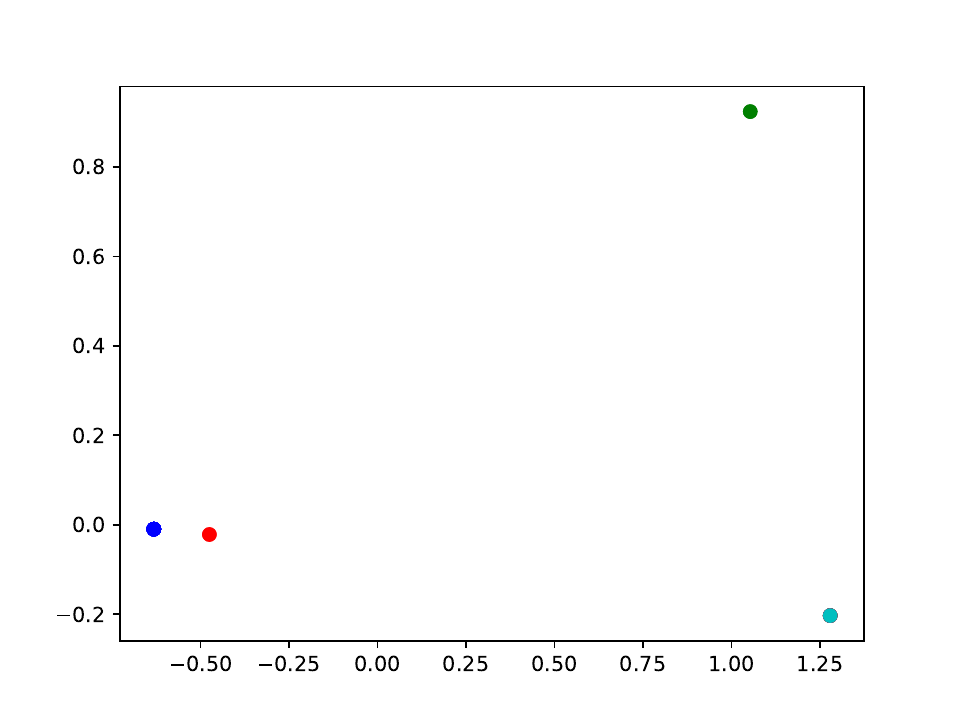}}
\hspace{-0.6cm}
\subfloat[GAS]{\includegraphics[width=0.27\textwidth]{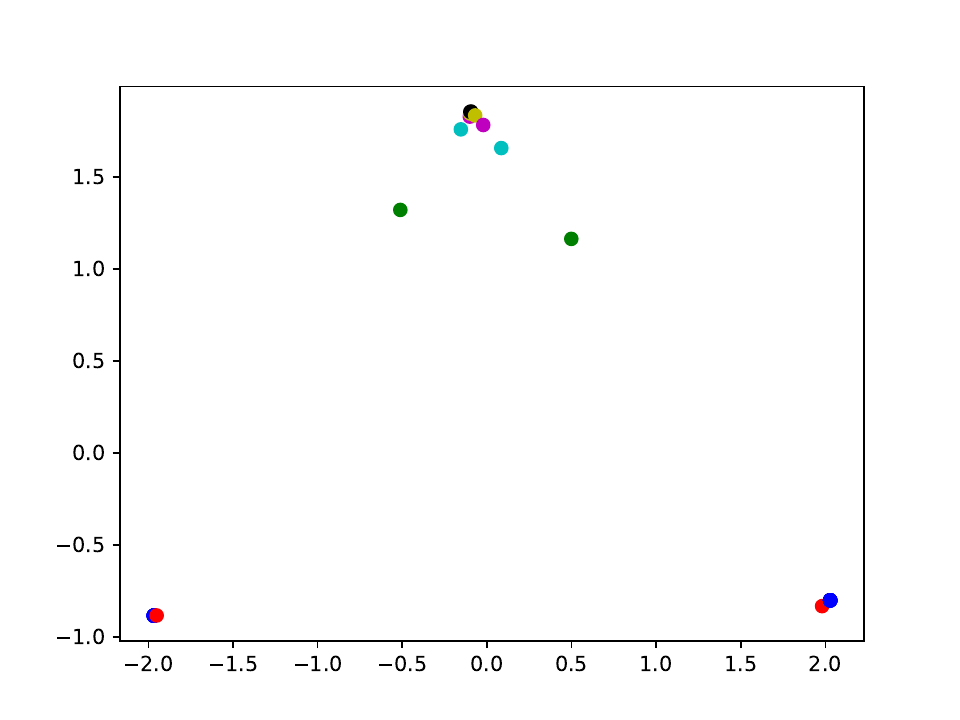}}
\hspace{-0.6cm}
\subfloat[RID$\varepsilon$Rs]{\includegraphics[width=0.27\textwidth]
{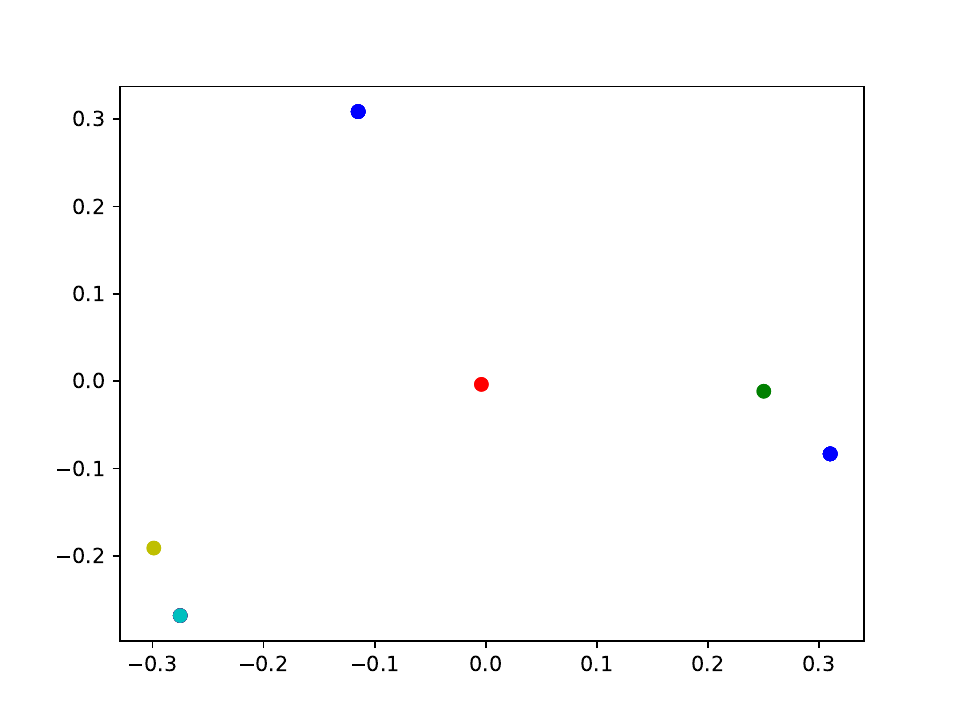}}
\caption{Barbell visualization with PCA.
}
\label{PCA}
\end{figure*}
\subsection{Visualization}
Network visualization is a fundamental task, particularly in scenarios where the roles of network nodes are predefined. The primary goal is to project these nodes into a two-dimensional space for visualization purposes. An effective embedding technique should position nodes with similar roles in close proximity within this reduced-dimensional space.

$\varepsilon$-BE, RID$\varepsilon$Rs, and Graphwave are not directly suitable for visualization. For this reason, we further reduce the embedding space of dimension $d$ into two dimensions using PCA and t-SNE \cite{van2008visualizing}.
The former is a deterministic approach based on matrix factorization.
The latter is a probabilistic  method based on a variation of Stochastic
Neighbor Embedding \cite{hinton2002stochastic}.
 
We consider the well-known barbell graph depicted in Figure \ref{barbell}, a standard benchmark in the literature for evaluating visualization tasks\cite{SEGK,GW,S2V}. It consists of two cliques connected by a chain of nodes, where each node is assigned a distinct role represented by different colors. The nodes within each clique share the same role, except for the red nodes, which are additionally connected to the chain. Conversely, the roles within the chain are specular. To demonstrate the efficacy of $\varepsilon$-BE, we compute it with $\varepsilon =  0$. 



Figure~\ref{TSNE} shows the visualizations by t-SNE reduction achieved by all the different methods.
We observe notable differences in the performance of different network embedding approaches. GAS and RID$\varepsilon$Rs exhibit poorer performance than the other methods, especially GAS, where the majority of nodes are clustered closely together in the embedding space, making it challenging to discern distinct groups.
In contrast, SEGK and Graphwave have superior performance by effectively distinguishing nodes within cliques. However, they tend to separate the red nodes into two separate regions within the embedding space. 
On the other hand, struc2vec and DRNE perform well, providing a good spatial organization. Interestingly, our approach yields results similar to struc2vec with a cleaner separation within the 2-dimensional space.

Considering the inherent probabilistic nature of t-SNE, we also considered a deterministic approach using PCA, shown in Figure~\ref{PCA}. In this case, nodes with the same encoding in the embedded space are mapped in the same point in the 2-dimensional space.
As a consequence, all nodes with the same role are mapped onto the same point for $\varepsilon$-BE. 
Graphwave, RID$\varepsilon$Rs, and SEGK show similar behavior, although they differentiate chain nodes less effectively. 
The other methods present a more sparse allocation without nodes that overlap perfectly. This is due to the fact that they rely on random walks and deep learning techniques. Consequently, it is unlikely to yield embedding vectors with identical components across these approaches. As a result, the application of PCA to visualize the embeddings generated by these methods yields less uniform results compared to other techniques.

Overall, for this case study  $\varepsilon$-BE approach emerges as the optimal strategy for accurate visualization using PCA.

\subsection{Classification}

The objective of classification is to train a model that can accurately assign appropriate roles to new nodes. 
We proceed by computing the embedding space for each network, followed by a 5-fold cross-validation to assess the accuracy of role prediction. 
To evaluate classification performance, we compute the average F1 score.

We consider different real-world networks. 
We started with the air traffic networks from Brazil, the United States, and Europe, presented in~\cite{S2V}; additionally, we use an actor co-occurrence network and the Film network from~\cite{RONE}. Air traffic networks are widely recognized within the network community and serve as standardized benchmarks for comparative evaluations \cite{S2V}. 


\begin{table}[t]
\centering
\resizebox{0.80\columnwidth}{!}{
 \begin{tabular}{l r r  r  r r }
 \toprule
    \emph{Network} & \multicolumn{1}{c}{$N$} & \multicolumn{1}{c}{\emph{d}} & \multicolumn{1}{c}{$\delta$} & \multicolumn{1}{c}{$\Delta$} &  \multicolumn{1}{c}{$\varepsilon_0$} \\
 \midrule
 Brazil airport & 131 & 38  & 1 &	3 &	 0  

\\
 Europe airport & 399 & 96 & 1 &	4 &	 0   
\\
 USA airport & 1190& 325 & 1 &	2 &	 0    
 \\
 actor& 7779& 201 & 2 &	8 &	 2   
\\
 Film & 27312& 177 & 6 &	30 &	 6    
\\
 \bottomrule
 \end{tabular}
 }
 \vspace*{3mm}
 \caption{ Parameter settings for the iterative $\varepsilon$-BE algorithm. }
 \label{param}
\end{table}

\begin{figure}[t]
\centering
\includegraphics[width=1\columnwidth]{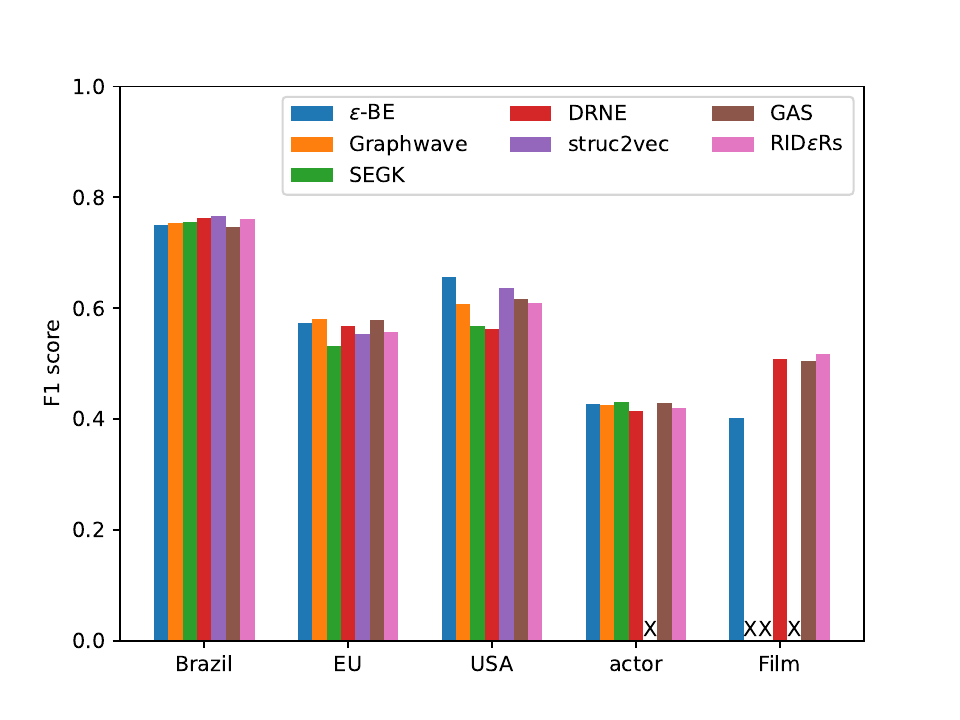}
\caption{Airport networks classification results. 
The bar indicated with `X' refers to methods that timed out (3\,h).}
\label{AirClass}
\end{figure}

All such networks come with ground truth labels that specify the role of each node. For each method, we computed the corresponding network embedding. 
We show in  Table~\ref{param}, together with the original network size $N$ and the embedding size $d$, the parameters utilized for computing the network embeddings with our approach.
We averaged the results over 50 5-fold cross-validations and reported them in Figure \ref{AirClass}.

In all the networks, $\varepsilon$-BE is able to achieve comparable results in the classification task. This is in line with recent findings claiming that there is currently no method that clearly outperforms the state of the art in this task~\cite{RONE}. 
The Film network proved challenging; however, three other methods (Graphwave, SEGK, and struct2vec) timed out in this case.

\begin{table*}[!ht]
\centering
\resizebox{0.9\linewidth}{!}{
 \begin{tabular}{l c c c c c c c c c c c c c c c}
 \toprule
      & \multicolumn{2}{c}{\emph{Brazil}} & \multicolumn{2}{c}{\emph{Europe}} & \multicolumn{2}{c}{\emph{United States}} & \multicolumn{2}{c}{\emph{Actor}} & \multicolumn{2}{c}{\emph{Film}}\\
  \midrule
 \emph{Method}  & EIG & BET & EIG & BET & EIG & BET & EIG & BET & EIG & BET\\
 \midrule
 $\varepsilon$-BE
  & \cellcolor{green!50}\textbf{8.47E-05}  &\cellcolor{green!50}\textbf{2.79E-02} 
 & \cellcolor{green!50}\textbf{1.23E-05} & \cellcolor{blue!25}7.56E-03
 & \cellcolor{green!50}\textbf{1.28E-05}    & \cellcolor{green!50}\textbf{4.12E-02}
 & \cellcolor{blue!25}9.93E-03  & \cellcolor{green!50}\textbf{1.73E-02}
 & \cellcolor{green!50}\textbf{2.30E-03} & \cellcolor{green!50}\textbf{8.10E-03}
 \\

 Graphwave & \cellcolor{blue!25}5.30E-04 &\cellcolor{blue!25}4.08E-02
& 8.51E-04      &1.17E-02
 &\cellcolor{blue!25}2.95E-03  &\cellcolor{blue!25}4.48E-02
 & 2.36E-02  & 5.87E-02
 & \emph{TO} & \emph{TO}\\

 SEGK &4.90E-03 &     6.17E-02
&3.38E-03    & 1.41E-02
 & 1.05E-02   &  5.12E-02
 & 2.29E-02   &   6.06E-02
 & \emph{TO}  & \emph{TO}\\

  DRNE & 3.41E-03    &4.81E-02
 & \cellcolor{blue!25}5.97E-04&    9.89E-03
 & 3.51E-03   &4.86E-02
 & \cellcolor{green!50}\textbf{5.58E-03}  & \cellcolor{blue!25}5.56E-02
  & 1.02E-02  & 1.30E-02\\

  struc2vec &3.68E-03   & 4.57E-02
  & 2.47E-03  & \cellcolor{green!50}\textbf{7.19E-03}
  & 5.66E-03 &   4.81E-02
  & \emph{TO}  & \emph{TO}
  & \emph{TO}  & \emph{TO}\\

    GAS &7.88E-02     &     4.19E-01
  & 1.56E-02&  2.69E-02
  & 1.62E-02&  6.21E-02
  & 1.87E-02 & 6.18E-01
   &1.04E-02  & 1.25E-02\\

   RID$\varepsilon$Rs &6.01E-03   &     5.03E-02
   & 5.09E-03 & 1.50E-02
    & 1.26E-02&  5.35E-02
    & 1.67E-02 &  6.01E-02
    & \cellcolor{blue!25}8.21E-3 & \cellcolor{blue!25}1.10E-02\\
   \bottomrule
 \end{tabular}
  }
 \vspace{2mm}
 \caption{Mean squared errors for the network regression task. EIG and BET stand for eigenvector and betweenness centrality. We highlight the best and second-best results in green and blue, respectively. \emph{TO}: timeout (3\,h)}
 \label{AirRegr}
\end{table*}

\subsection{Regression}
In regression, the objective is to predict a continuous or numerical value based on input features.
Following~\cite{DRNE}, we define the target score for regression by incorporating two well-recognized centrality measures: eigenvector and betweenness centrality.
The idea is to use the encoding of each node in the embedding space as a predictor for the centrality measures in the original network. 

The goal is to compare methods with respect to their ability to preserve the centrality measures in the lower dimensional space.
This is significant because the centrality measures have established correlations with the fundamental notions of role and position within a network \cite{DRNE}.

We train a linear regression allocating 20\% of the nodes as the designated test set and employing the remainder for training. 
We use the trained linear regression model to predict the centrality score for each node in the test set. In line with~\cite{DRNE}, we employ the Mean Square Error (MSE) metric to quantify the discrepancy between the predicted and the true centrality. The former corresponds to the centrality predicted for each node in the test set. The latter corresponds to the centrality of these nodes in the real networks where their encoding corresponds to the rows of the adjacency matrix $A$.
To normalize for the different scales inherent to centrality measures, the resultant value is divided by the average value.

The analysis results are shown in Table~\ref{AirRegr}.
$\varepsilon$-BE  consistently exhibits optimal performance in eigenvector and betweenness centrality across nearly all network instances. This outcome is aligned with the intrinsic properties of the BE reduction methodology, which effectively preserves eigenvector centrality, Katz centrality, and PageRank centrality, as reported in the work ~\cite{Isola}. Although RID$\varepsilon$Rs should inherit this property too, since it is based on approximate equitable partitions, it demonstrated good performance (second-best) for eigenvalue centrality only in the Film network, while it was outperformed by the competition in the other networks.  
Finally, we observe that $\varepsilon$-BE is highly competitive since it is the method with always the best or second-best performance. Struc2vec, Graphwave, and DRNE display good performances but not comparable with our approach. Additionally, some of them timed out when analyzing the larger networks. 

\begin{table}[t]
\centering
\resizebox{\columnwidth}{!}{
 \begin{tabular}{l c  c  c c c }
 \toprule
    & \multicolumn{5}{c}{\emph{Networks}} \\
    \cmidrule(l){2-6}
    \emph{Method} & \multicolumn{1}{c}{\emph{Brazil}} & \multicolumn{1}{c}{\emph{USA }} & \multicolumn{1}{c}{\emph{BlogCatalog}} &  \multicolumn{1}{c}{\emph{Brightkite}} 
    &  \multicolumn{1}{c}{\emph{GitHub}}\\ 
    \cmidrule(l){2-6}
    & $n = 131$ & $1190$ & $10312$ & $58689$ & $177316$  \\
    \cmidrule(l){2-6} 
    
 \midrule
 $\varepsilon$-BE & \cellcolor{green!50}\textbf{2.94E-1} &	\cellcolor{green!50}\textbf{6.49E-1} &	 \cellcolor{green!50}\textbf{6.24E+0} & \cellcolor{green!50}\textbf{7.68E+1} & \cellcolor{green!50}\textbf{4.81E+2}

\\
 Graphwave & 2.52E+0	& 1.03E+1	&1.47E+3 & \emph{TO} & \emph{TO}
\\
 SEGK &1.92E+1	&4.58E+3&	\emph{TO} & \emph{TO} & \emph{TO}
 \\
 DRNE & 1.07E+1	&2.50E+1	& \cellcolor{blue!25}1.96E+2  & 1.26E+3 & \emph{TO}
\\
 struc2vec &1.83E+1	&9.93E+3&	\emph{TO} & \emph{TO} & \emph{TO}
\\
 GAS & \cellcolor{blue!25}2.30E+0	& \cellcolor{blue!25}7.95E+0&	4.90E+2 & \cellcolor{blue!25}4.98E+2 & \cellcolor{blue!25}2.65E+3
\\
 RID$\varepsilon$Rs &3.51E+1	&2.73E+2&	3.96E+2 & 7.38E+3 & \emph{TO}
\\
 \bottomrule
 \end{tabular}
 }
 \vspace*{3mm}
 \caption{Network embedding running times (in s). \emph{TO}: timeout (3\,h).}
 \label{RT}
\end{table} 

\subsection{Scalability}

We empirically analyzed the scalability of $\varepsilon$-BE by comparing its running times against the other methods on networks of increasing size. For this, we also considered three larger networks, BlogCatalog \cite{BlogCat}, Brightkite \cite{bright}, and GitHub \cite{GitH}. 

Results are reported in Table~\ref{RT}, showing that $\varepsilon$-BE consistently demonstrated superior runtime performance. Specifically, SEGK and struc2vec timed out with the BlogCatalog network, while Graphwave timed out with the Brightkite network. On the GitHub network, DRNE encountered a runtime error, which stopped the embedding process prematurely. 
RID$\varepsilon$Rs timed out only with the GitHub network, but its running times are at least two orders of magnitude larger than $\varepsilon$-BE (while it performed significantly better than $\varepsilon$-BE only in the classification task of the Film network).  The GitHub network could be effectively handled only by $\varepsilon$-BE and GAS. Although GAS exhibited competitive running times (which are, however, one order of magnitude slower), its performance was inferior, except for the classification of the Film network.

\section{Conclusion} 

Structural network embeddings are useful for tasks that exploit properties of similarities of the roles of the nodes in a network. Our network embedding is based on $\varepsilon$-BE, which partitions the nodes of a network into approximately equitable blocks using an efficient partition refinement algorithm embedded in an iterative framework that controls the impact of the tolerance parameter $\varepsilon$. The experimental evaluation on visualization, classification, and regression tasks demonstrate that $\varepsilon$-BE consistently achieves comparable or superior performance compared to existing methods at shorter running times compared to other state-of-the-art methods. 

While $\varepsilon$-BE currently addresses undirected binary networks, it appears to be extensible to directed and weighted networks by suitably accommodating the notion of equitable partition for both in- and out-neighbors. 
Additionally, we plan to enhance $\varepsilon$-BE by considering different initial partitions based on the problem context. While this study adopts an initial partition where all nodes fall into a single block to ensure a fair comparison with related works, incorporating problem-specific initial partitions may lead to improved performance.

\bibliographystyle{IEEEtran}
\bibliography{ICDM}

\end{document}